\documentclass[12pt]{article}

\usepackage{axodraw}

\makeatletter

\def\paragraph{\@startsection{paragraph}{4}{\z@}{+2.00ex plus
 +1ex minus +.2ex}{1.5ex plus .2ex}{\it\normalsize}}

\def\section{\@startsection {section}{1}{\z@}{+3.0ex plus +1ex minus
  +.2ex}{2.3ex plus .2ex}{\normalsize\bf\boldmath}}
\def\subsection{\@startsection{subsection}{2}{\z@}{+2.5ex plus +1ex
minus +.2ex}{1.5ex plus .2ex}{\normalsize\bf\boldmath}}
\def\subsubsection{\@startsection{subsubsection}{3}{\z@}{+3.25ex plus
 +1ex minus +.2ex}{1.5ex plus .2ex}{\normalsize\it}}

\expandafter\ifx\csname mathrm\endcsname\relax\def\mathrm#1{{\rm #1}}\fi

\newcounter{saveeqn}

\@addtoreset{equation}{section}

\newcount\@tempcntc
\def\@citex[#1]#2{\if@filesw\immediate\write\@auxout{\string\citation{#2}}\fi
  \@tempcnta\z@\@tempcntb\m@ne\def\@citea{}\@cite{\@for\@citeb:=#2\do
    {\@ifundefined
       {b@\@citeb}{\@citeo\@tempcntb\m@ne\@citea
        \def\@citea{,\penalty\@m\ }{\bf ?}\@warning
       {Citation `\@citeb' on page \thepage \space undefined}}%
    {\setbox\z@\hbox{\global\@tempcntc0\csname
b@\@citeb\endcsname\relax}%
     \ifnum\@tempcntc=\z@ \@citeo\@tempcntb\m@ne
       \@citea\def\@citea{,\penalty\@m}
       \hbox{\csname b@\@citeb\endcsname}%
     \else
      \advance\@tempcntb\@ne
      \ifnum\@tempcntb=\@tempcntc
      \else\advance\@tempcntb\m@ne\@citeo
      \@tempcnta\@tempcntc\@tempcntb\@tempcntc\fi\fi}}\@citeo}{#1}}

\def\@citeo{\ifnum\@tempcnta>\@tempcntb\else\@citea
  \def\@citea{,\penalty\@m}%
  \ifnum\@tempcnta=\@tempcntb\the\@tempcnta\else
   {\advance\@tempcnta\@ne\ifnum\@tempcnta=\@tempcntb \else
\def\@citea{--}\fi
    \advance\@tempcnta\m@ne\the\@tempcnta\@citea\the\@tempcntb}\fi\fi}

\newcommand{\lsim}
{\mathrel{\raisebox{-.3em}{$\stackrel{\displaystyle <}{\sim}$}}}
\newcommand{\gsim}
{\mathrel{\raisebox{-.3em}{$\stackrel{\displaystyle >}{\sim}$}}}
\def\asymp#1%
{\mathrel{\raisebox{-.4em}{$\widetilde{\scriptstyle #1}$}}}

\def\Nequal#1%
{\mathrel{\raisebox{-.5em}{$\stackrel{=}{\scriptstyle\rm#1}$}}}
\newcommand{\dsl}[1]{\not \hspace{-0.7mm}#1}
\def\dsl{\mathpalette\make@slash}
\def\make@slash#1#2{\setbox\z@\hbox{$#1#2$}%
  \hbox to 0pt{\hss$#1/$\hss\kern-\wd0}\box0}

\def\beq{\begin{equation}}
\def\eeq{\end{equation}}
\def\beqar{\begin{eqnarray}}
\def\eeqar{\end{eqnarray}}
\def\barr#1{\begin{array}{#1}}
\def\earr{\end{array}}
\def\bfi{\begin{figure}}
\def\efi{\end{figure}}
\def\btab{\begin{table}}
\def\etab{\end{table}}
\def\bce{\begin{center}}
\def\ece{\end{center}}

\def\text{\textstyle}

\def\ga{\gamma}
\def\de{\delta}

\def\si{\sigma}

\def\reffi#1{\mbox{Figure~\ref{#1}}}

\def\refse#1{\mbox{Section~\ref{#1}}}

\def\citere#1{\mbox{Ref.~\cite{#1}}}
\def\citeres#1{\mbox{Refs.~\cite{#1}}}

\newcommand{\TeV}{\unskip\,\mathrm{TeV}}
\newcommand{\GeV}{\unskip\,\mathrm{GeV}}
\newcommand{\MeV}{\unskip\,\mathrm{MeV}}

\newcommand{\fb}{\unskip\,\mathrm{fb}}

\newcommand{\ri}{{\mathrm{i}}}
\newcommand{\rd}{{\mathrm{d}}}

\newcommand{\Oa}{\mathswitch{{\cal{O}}(\alpha)}}

\def\mathswitchr#1{\relax\ifmmode{\mathrm{#1}}\else$\mathrm{#1}$\fi}

\newcommand{\PW}{\mathswitchr W}

\newcommand{\PZ}{\mathswitchr Z}

\newcommand{\Pg}{\mathswitchr g}

\newcommand{\Pe}{\mathswitchr e}

\newcommand{\Pnebar}{\mathswitch \bar\nu_{\mathrm{e}}}
\newcommand{\Pd}{\mathswitchr d}
\newcommand{\Pdbar}{\bar{\mathswitchr d}}
\newcommand{\Pu}{\mathswitchr u}

\newcommand{\Ps}{\mathswitchr s}

\newcommand{\Pc}{\mathswitchr c}

\newcommand{\Pep}{\mathswitchr {e^+}}
\newcommand{\Pem}{\mathswitchr {e^-}}

\def\mathswitch#1{\relax\ifmmode#1\else$#1$\fi}

\newcommand{\MW}{\mathswitch {M_\PW}}

\newcommand{\MZ}{\mathswitch {M_\PZ}}

\newcommand{\Me}{\mathswitch {m_\Pe}}

\newcommand{\GW}{\Gamma_{\PW}}

\def\solid{\raise.9mm\hbox{\protect\rule{1.1cm}{.2mm}}}
\def\dash{\raise.9mm\hbox{\protect\rule{2mm}{.2mm}}\hspace*{1mm}}

\def\ie{i.e.\ }

\newcommand{\eeWWffff}{\Pep\Pem\to\PW\PW\to 4f}
\newcommand{\eeWWffffg}{\eeWWffff\gamma}
\newcommand{\eeffff}{\Pep\Pem\to 4f}

\def\draftdate{\relax}
\def\mda{\relax}
\def\mua{\relax}
\def\mla{\relax}
\def\draft{
\def\thtystars{******************************}
\def\sixtystars{\thtystars\thtystars}
\typeout{}
\typeout{\sixtystars**}
\typeout{* Draft mode!
         For final version remove \protect\draft\space in source file *}
\typeout{\sixtystars**}
\typeout{}
\def\draftdate{\today}
\def\mua{\marginpar[\boldmath\hfil$\uparrow$]%
                   {\boldmath$\uparrow$\hfil}%
                    \typeout{marginpar: $\uparrow$}\ignorespaces}
\def\mda{\marginpar[\boldmath\hfil$\downarrow$]%
                   {\boldmath$\downarrow$\hfil}%
                    \typeout{marginpar: $\downarrow$}\ignorespaces}
\def\mla{\marginpar[\boldmath\hfil$\rightarrow$]%
                   {\boldmath$\leftarrow $\hfil}%
                    \typeout{marginpar: $\leftrightarrow$}\ignorespaces}
\def\Mua{\marginpar[\boldmath\hfil$\Uparrow$]%
                   {\boldmath$\Uparrow$\hfil}%
                    \typeout{marginpar: $\uparrow$}\ignorespaces}
\def\Mda{\marginpar[\boldmath\hfil$\Downarrow$]%
                   {\boldmath$\Downarrow$\hfil}%
                    \typeout{marginpar: $\downarrow$}\ignorespaces}
\def\Mla{\marginpar[\boldmath\hfil$\Rightarrow$]%
                   {\boldmath$\Leftarrow $\hfil}%
                    \typeout{marginpar: $\leftrightarrow$}\ignorespaces}
\overfullrule 5pt
\oddsidemargin -15mm
\marginparwidth 29mm
}

\def\stars{\strut\leaders\hbox{*}\hfill\strut}
\def\starline{\hfil\strut\hfil\hbox to \textwidth {\stars}\hfil}

\begin{document}
\thispagestyle{empty}
\def\thefootnote{\fnsymbol{footnote}}
\setcounter{footnote}{1}
\null
\draftdate\hfill BI-TP 99/23 \\
\strut\hfill PSI-PR-99-23\\
\strut\hfill hep-ph/9909363
\vfill
\begin{center}
{\Large \bf\boldmath
Radiative corrections to $\Pep\Pem\to\PW\PW\to4\mbox{ fermions}$%
\footnote{To appear in the Proceedings of the International Workshop on
Linear Colliders, Sitges, Barcelona, Spain, April 28 -- May 5, 1999}
\par} \vskip 2.5em
\vspace{1cm}

{\large
{\sc A.\ Denner$^1$, S.\ Dittmaier$^2$, M. Roth$^{1,3}$ and
D.\ Wackeroth$^1$} } \\[1cm]

$^1$ {\it Paul-Scherrer-Institut, W\"urenlingen und Villigen\\
CH--5232 Villigen PSI, Switzerland} \\[0.5cm]

$^2$ {\it Theoretische Physik, Universit\"at Bielefeld \\
D-33615 Bielefeld, Germany}
\\[0.5cm]

$^3$ {\it Institut f\"ur Theoretische Physik, ETH-H\"onggerberg\\
CH--8093 Z\"urich, Switzerland}
\par \vskip 1em
\end{center}\par
\vskip 2cm {\bf Abstract:} \par
The structure of the double-pole approximation for the
${\cal O}(\alpha)$ corrections to $\Pep\Pem\to\PW\PW\to 4\,$fermions is
described, and some results are presented. Moreover, results on full
tree-level predictions for $\Pep\Pem\to 4\,$fermions$+\gamma$ are given.
\par
\vskip 1cm
\noindent
September 1999
\null
\setcounter{page}{0}
\clearpage
\def\thefootnote{\arabic{footnote}}
\setcounter{footnote}{0}

\section{Introduction}

The investigation of the reactions 
$\Pep\Pem\to\PW\PW\to 4\,$fermions$\,(+\gamma)$ at future high-energy and
high-luminosity linear colliders is very important, since it provides us
with precise information about the W-boson mass $\MW$ and the
gauge-boson self-interactions. 
The most promising methods for the
determination of $\MW$ are the cross-section measurement at the W-pair
threshold and the reconstruction of the invariant masses of the W~bosons
at any energies. The experimental accuracies of these two measurements are 
expected to be of the order of $6\MeV$ \cite{wi99} and $15\MeV$ \cite{ecfa} 
for TESLA, respectively, which should be compared with the expected
accuray of $30\MeV$ at LEP2. While the triple gauge-boson couplings are 
probed at LEP2 at the level of 10\%, future $\Pep\Pem$ linear colliders can 
even exceed the per-cent level \cite{ecfa}. At future colliders it will also 
be possible to derive significant bounds on quartic gauge-boson couplings 
by inspecting W-pair production in association with a hard 
photon \cite{be92}; even at LEP2, where the statistics for such events is
poor, first bounds on quartic couplings can be derived \cite{opal99}.

The described physical goals can only be reached if precise predictions
for the processes $\Pep\Pem\to\PW\PW\to 4f(+\gamma)$ are known.
Assuming an integrated luminosity of the order of
$10^2\fb^{-1}$ leads to about $10^6$ pairs of W~bosons. This means that 
physical observables should be known at the level of some $0.1\%$. 
High-precision calculations for four-fermion production are,
however, complicated for various reasons. At the aimed accuracy, a pure
on-shell approximation for the W~bosons is not acceptable, i.e.\ the
W~bosons have to be treated as resonances. Since the description of
resonances necessarily goes beyond a fixed-order calculation in
perturbation theory, problems with gauge invariance occur.
Discussions of this issue can be found in \citeres{lep2repWcs,flscheme}.
A second complication arises from the need to take into account 
electroweak radiative corrections of ${\cal O}(\alpha)$ beyond 
the universal corrections.
The full treatment of the processes $\Pep\Pem\to 4f$ at the
one-loop level is of enormous complexity and involves severe theoretical
problems with gauge invariance; up to now such results do not exist.

Here we summarize recent progress concerning an approximate approach to
include ${\cal O}(\alpha)$ corrections to $\Pep\Pem\to\PW\PW\to 4f$. The
approximation is based on the idea to correct only those pieces of the
transition matrix elements that are enhanced by two W-boson resonances
and is therefore called 
{\it double-pole approximation} (DPA). Corrections of ${\cal O}(\alpha)$
to contributions that involve at most one resonant W~boson are
of the order of $(\alpha/\pi)\times(\GW/\MW)\times\log(\cdots)\lsim 0.1\%$, 
which thus is a measure for the intrinsic uncertainty of the DPA.
Corrections induced by real photon emission may be treated accordingly,
but full tree-level predictions for $\Pep\Pem\to 4f+\gamma$ have already 
been presented for selected final states in \citeres{ee4fa0} and for all
final states in \citere{ee4fa}. The most important results of 
\citere{ee4fa} are reviewed below.
\looseness -1

\section{\boldmath{Full tree-level predictions for $\Pep\Pem\to 4f+\gamma$}}

The processes $\Pep\Pem\to 4f+\gamma$ do not only yield important 
corrections to $\Pep\Pem\to 4f$, they are also
interesting in their own right, since they involve both triple and quartic 
gauge-boson couplings. 

Most of the existing work on hard-photon radiation in W-pair production
is based on the approximation of stable W~bosons (see \citere{brrceeww}
and references in \citeres{lep2repWcs,ee4fa}). 
A first step of including the off-shellness of W~bosons in
$\Pep\Pem\to\PW\PW\to 4f+\gamma$ was done in \citere{ae91}, where only
photon emission from diagrams with two resonant W~bosons was taken
into account. 
However, it is desirable to have a full lowest-order calculation
for $\Pep\Pem\to 4f+\gamma$ for two reasons. 
As described in \refse{se:DPA},
the definition of the DPA for $\Pep\Pem\to\PW\PW\to 4f+\gamma$ is
non-trivial so that possible versions of the DPA should be carefully
compared to the full result. Secondly, one expects a similar impact of
off-shell effects as in the case without photon, where so-called
background diagrams (diagrams with at most one resonant W~boson) can
reach a significant fraction of the full cross section.
Therefore, they should be included at least in predictions for 
detectable photons.
In the following we briefly summarize some results of \citere{ee4fa},
where an event generator for all final states
$4f+\gamma$ with massless fermions is described.

In the event generator of \citere{ee4fa} 
different schemes for treating gauge-boson widths are implemented. 
A comparison of results obtained by different ways of introducing these
decay widths is useful in order to get information about the size of
gauge-invariance-breaking effects, which are present in some
finite-width schemes.
\begin{table}
\begin{center}
{\begin{tabular}{|c|c|r@{}l|r@{}l|r@{}l|r@{}l|}
\hline
\multicolumn{1}{|c|}{$\si/\fb$} &
\multicolumn{1}{r|}{$\sqrt{s}=$} & 
\multicolumn{2}{c|}{$189\GeV$} & 
\multicolumn{2}{c|}{$500\GeV$} &
\multicolumn{2}{c|}{$2\TeV$} & 
\multicolumn{2}{c|}{$10\TeV$}
\\\hline\hline
& constant width
&$    224.0 $&$( 4)$
&$     83.4 $&$( 3)$
&$     6.98 $&$( 5)$
&$    0.457 $&$( 6)$
\\\cline{2-10}
$\Pep \Pem \to \Pu\, \Pdbar\, \mu^- \bar{\nu}_\mu \,\ga$
& running width
&$    224.6 $&$( 4)$
&$     84.2 $&$( 3)$
&$     19.2 $&$( 1)$
&$      368 $&$( 6)$
\\\cline{2-10}
& complex mass          
&$    223.9 $&$( 4)$
&$     83.3 $&$( 3)$
&$     6.98 $&$( 5)$
&$    0.460 $&$( 6)$
\\\hline\hline
& constant width
&$    230.0 $&$( 4)$
&$    136.5 $&$( 5)$
&$     84.0 $&$( 7)$
&$     16.8 $&$( 5)$
\\\cline{2-10}
$\Pep \Pem \to \Pu\, \Pdbar\, \Pe^- \Pnebar \,\ga $
& running width
&$    230.6 $&$( 4)$
&$    137.3 $&$( 5)$
&$     95.7 $&$( 7)$
&$      379 $&$( 6)$
\\\cline{2-10}
& complex mass          
&$    229.9 $&$( 4)$
&$    136.4 $&$( 5)$
&$     84.1 $&$( 6)$
&$     16.8 $&$( 5)$
\\\hline
\end{tabular}}
\end{center}
\caption[]{Comparison of different width schemes for several processes 
(taken from \citere{ee4fa})}
\label{tab:ee4fawidth}
\end{table}
Table~\ref{tab:ee4fawidth} contains some 
results on the total cross section for 
two semi-leptonic four-fermion final states and a photon, evaluated 
with different finite-width treatments.
Similar to the case without photon emission, the SU(2)-breaking effects
induced by a running width render the predictions totally wrong in the
TeV range. For a constant width such effects are suppressed, as can be
seen from a comparison with the results of the complex-mass scheme,
which exactly preserves gauge invariance.

\begin{figure}
\centerline{
\setlength{\unitlength}{1cm}
\begin{picture}(6,5.6)
\put(0,0){\includegraphics{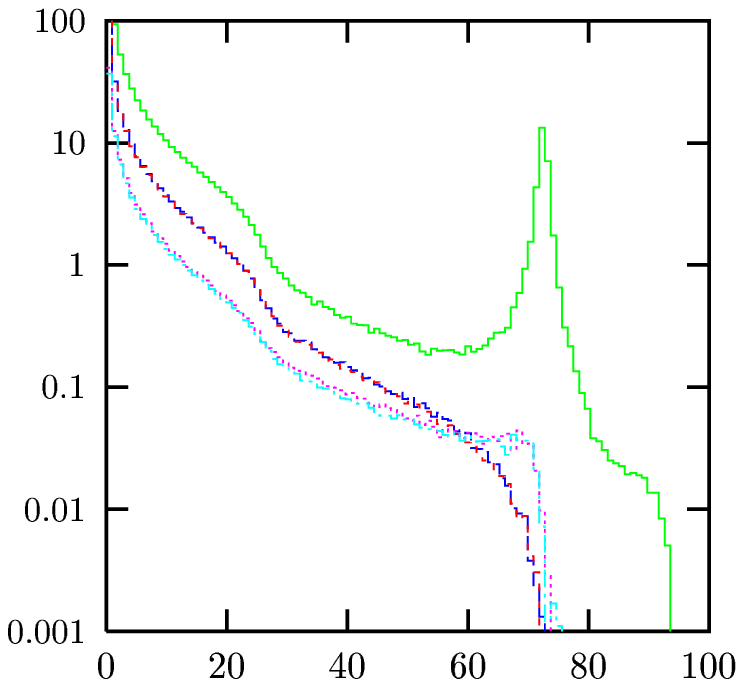}}
\put(1.5,1.4){\footnotesize $\sqrt{s}=189\GeV$}
\put(3.0,-0.7){\makebox(1,1)[cc]{{\small $E_\ga/\GeV$}}}
\end{picture}
\begin{picture}(6,5.6)
\put(0,0){\includegraphics{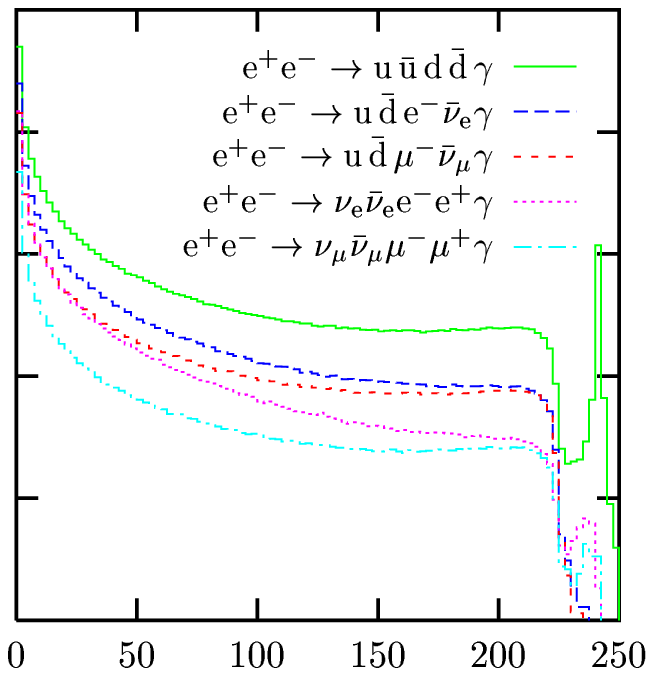}}
\put(1.1,1.4){\footnotesize $\sqrt{s}=500\GeV$}
\put(2.5,-0.7){\makebox(1,1)[cc]{{\small $E_\ga/\GeV$}}}
\end{picture} }
\caption[]{Photon-energy spectra 
$(\rd\si/\rd E_\ga)/(\fb/\GeV)$ for several processes 
(taken from \citere{ee4fa})}
\label{fi:photonspectra}
\end{figure}
\begin{figure}
\centerline{
\begin{picture}(150,100)(-5,0)
\SetScale{.8}
\ArrowLine(0,10)(40,50)
\ArrowLine(40,70)(0,110)
\Photon(50,60)(150,60){2}{11}
\Photon(50,63)(120,90){2}{8}
\Photon(50,57)(120,30){2}{8}
\Vertex(120,90){2}
\Vertex(120,30){2}
\ArrowLine(120,90)(150,110)
\ArrowLine(150,70)(120,90)
\ArrowLine(120,30)(150,50)
\ArrowLine(150,10)(120,30)
\GCirc(50,60){15}{0}
\put(75,18){\makebox(1,1)[c]{$V_2$}}
\put(75,78){\makebox(1,1)[c]{$V_1$}}
\put(128,50){\makebox(1,1)[c]{$\gamma$}}
\Text(-5,100)[rt]{a)}
\SetScale{1}
\end{picture}
\begin{picture}(160,100)
\SetScale{.8}
\ArrowLine(0,10)(40,50)
\ArrowLine(40,70)(0,110)
\Photon(50,63)(150,100){2}{12}
\Photon(50,57)(110,30){2}{8}
\Vertex(110,30){2}
\ArrowLine(110,30)(140,50)\ArrowLine(140,50)(170,70)
\ArrowLine(150,10)(110,30)
\Vertex(140,50){2}
\Photon(140,50)(170,30){2}{4}
\Vertex(170,30){2}
\ArrowLine(170,30)(200,50)
\ArrowLine(200,10)(170,30)
\GCirc(50,60){15}{0}
\put(65,24){\makebox(1,1)[c]{$Z$}}
\put(133,38){\makebox(1,1)[c]{$V_3$}}
\put(128,80){\makebox(1,1)[c]{$\gamma$}}
\Text(-5,100)[rt]{b)}
\SetScale{1}
\end{picture}
}
\vspace*{-1em}
\caption[]{Diagrams for important subprocesses in $4f+\gamma$ production
($V_1,V_2=\PW,\PZ,\ga$, $V_3=\ga,\Pg$) }
\label{fi:resonance}
\end{figure}
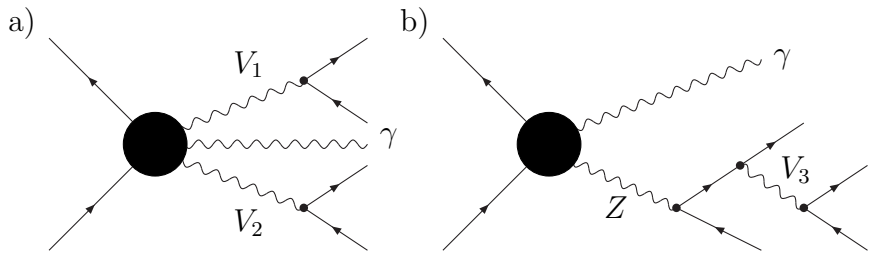

Figure~\ref{fi:photonspectra} shows the photon-energy spectra for some
typical four-fermion final states that correspond to $\PW\PW\gamma$
production. Apart from the usual soft-photon pole, the spectra contain
several threshold and peaking structures that are caused by photon
emission from the initial state. The two relevant classes of diagrams
are illustrated in \reffi{fi:resonance}.
Diagrams with
the structure of \reffi{fi:resonance}a correspond to
triple-gauge-boson-production subprocesses and yield dominant
contributions as long as the two virtual gauge bosons can become
simultaneously resonant. For instance, $\PW\PW\gamma$ production is
dominant for $E_\gamma < 26.3\GeV$ ($224\GeV$) for a CM
energy of $189\GeV$ ($500\GeV$). The diagrams of \reffi{fi:resonance}b
correspond to $\gamma\PZ$ production with a subsequent four-particle
decay of the resonant Z~boson mediated by a soft photon or gluon $V_3$.
Owing to the two-particle kinematics of $\gamma\PZ$ production such
contributions lead to peak structures around a fixed value of $E_\gamma$,
which is located at $72.5\GeV$ ($242\GeV$) for a CM energy of $189\GeV$
($500\GeV$). 

Table \ref{tab:ee4fawidth} and \reffi{fi:photonspectra} illustrate the
effect of background diagrams, since final states that are related by
the interchange of muon and electrons differ only by background
diagrams. 
While the impact of background diagrams is of the order of some per cent
for CM energies around $200\GeV$, there is a large effect of background 
contributions already at $500\GeV$.
The main effect is due to forward-scattered $\Pe^\pm$, which is familiar
from the results on $\Pep\Pem\to 4f$. More numerical results for
$\Pep\Pem\to 4f+\gamma$ can be found in \citere{ee4fa}.

\section{Electroweak radiative corrections}

\subsection{Relevance of electroweak corrections}

\begin{sloppypar}
Present-day Monte Carlo generators
for off-shell W-pair production (see e.g.\ \citere{lep2repWevgen})
typically include only universal electroweak ${\cal O}(\alpha)$ corrections%
\footnote{The QCD corrections for hadronic final states are discussed 
in \citere{qcdcorr}.},
such as the running of the electromagnetic coupling, $\alpha(q^2)$, 
leading corrections entering via the $\rho$-parameter,
the Coulomb singularity \cite{coul}, which is important near threshold,
and mass-singular logarithms 
$\alpha\ln(\Me^2/Q^2)$ from initial-state radiation. 
In leading logarithmic approximation, 
the scale $Q^2$ is not determined and has to be set to
a typical scale for the process; in the following we take $Q^2=s$.
\end{sloppypar}

The size of the neglected ${\cal O}(\alpha)$ contributions is
estimated by inspecting on-shell W-pair production, for which the exact 
${\cal O}(\alpha)$ correction and the leading contributions were
given in \citeres{vrceeww} and \cite{bo92}, respectively.
Table \ref{tab:nlRCs} shows the difference between an ``improved
Born approximation'' $\de_{\mathrm{IBA}}$, which is based on the 
above-mentioned universal corrections, and the corresponding full
${\cal O}(\alpha)$ correction $\de$ to the Born cross-section integrated over
the W-production angle $\theta$ for some centre-of-mass (CM) energies 
$\sqrt{s}$.
\begin{table}
\centerline{
\begin{tabular}{|c||c||c|c|c|c|c|c|}
\hline
$\theta$ range & $\sqrt{s}/\GeV$ &
161 & 175 & 200 & 500 & 1000 & 2000 \\
\hline\hline
$0^\circ$$<$$\theta$$<$$180^\circ$ & 
$(\de_{\mathrm{IBA}}-\de)/\%$ 
& 1.5 & 1.3 & 1.5 & 3.7 & 6.0 & 9.3 \\
\cline{1-1} \cline{3-8}
$10^\circ$$<$$\theta$$<$$170^\circ$ &&
1.5 & 1.3 & 1.5 & 4.7 & 11 & 22 \\
\hline
\end{tabular}
}
\caption{Size of ``non-leading'' corrections to on-shell W-pair
production ($\de_{\mathrm{IBA}}$ and $\de$ include only soft-photon
emission)}
\label{tab:nlRCs}
\end{table}
More details and results can be found in \citeres{lep2repWcs,crad96}.
The quantity $\de_{\mathrm{IBA}}-\de$ corresponds to the neglected
non-leading corrections and amounts to $\sim 1$--2\% for LEP2 energies,
but to $\sim 10$--20\% in the TeV range. Thus, in view of the desired
accuracy of some $0.1\%$, the inclusion of non-leading corrections is
indispensable. 
The large contributions in $\de_{\mathrm{IBA}}-\de$ at high energies are
due to double-logarithmic 
terms such as $\alpha\ln^2(s/\MW^2)$, which arise from vertex 
and box corrections and can be read off from the high-energy 
expansion \cite{be93} of the virtual and soft-photonic ${\cal O}(\alpha)$ 
corrections.

\subsection{Photon radiation and W line shape}
\label{se:Wlineshape}

A thorough description of real-photon emission is of particular
importance for the realistic prediction of the W~line shape, which
is the basic observable for the reconstruction of the W-boson mass from
the W-decay products. This fact can be easily understood by comparing the
impact of photon radiation on the line shape of the W~boson with the one
of the Z~boson, observed in $\Pep\Pem\to\PZ\to f\bar f$ at LEP1 and the
SLC (see also \citere{bbc98c}).

The Z~line shape is defined as a function of $s$, which is fully 
determined by the initial state, by the cross section $\sigma(s)$.
Photon radiation from the initial state effectively reduces the value of 
$s$ available for the production of 
the Z boson so that $\sigma(s)$ also receives
resonant contributions for $s>\MZ^2$, induced by this {\it radiative
return} to the Z~resonance and known as {\it radiative tail}. Final-state 
radiation is not enhanced by
such kinematical effects, thus yielding moderate corrections.

The W~line shape is reconstructed from the kinematical variables in the
final state. More precisely, it is defined by the distributions
$\rd\sigma/\rd M_\pm^2$, where $M_\pm^2$ are the reconstructed invariant
masses of the $\PW^\pm$~bosons. We now consider the fermion pair 
$f_1(k_1) \bar f_2(k_2)$ produced by a nearly resonant W~boson with momentum
$k_+$, i.e.\ $k_+^2\sim\MW^2$.
In this case, photon radiation from the final state decreases the
invariant mass of this fermion pair, i.e.\ 
$(k_1+k_2)^2<k_+^2=(k_1+k_2+k_\gamma)^2$,
while initial-state radiation leads to 
$(k_1+k_2)^2=k_+^2<(k_1+k_2+k_\gamma)^2$.
Thus, a consistent identification of $M_+^2=(k_1+k_2)^2$ also leads to a
radiative tail, but now induced by final-state radiation and for 
$M_+^2<\MW^2$. However, such an identification is experimentally not
possible for almost all cases%
\footnote{Semi-leptonic final states with a muon may be an exception,
where $(k_\mu+k_{\nu_\mu})^2$ could be determined from all detected
final-state particles other than the muon.},
since nearly collinear and soft photons in the final state cannot be 
separated from the outgoing fermions (except for muons). A realistic
definition of $M_\pm^2$ necessarily depends on the details of the
experimental treatment of photons in the final state, underlining the
importance of a careful investigation of the W~line shape in the
presence of photon radiation.

\subsection{Features of the double-pole approximation}
\label{se:DPA}

Fortunately, the full off-shell calculation for the processes
$\Pep\Pem\to\PW\PW\to 4f$ in ${\cal O}(\alpha)$ is not needed for
most applications. Sufficiently above the W-pair threshold 
a good approximation can be obtained by taking into account
only the doubly-resonant part of the amplitude 
\beq
\label{eq:Mstruc}
{\cal M} =
\underbrace{\frac{R_{+-}(k_+^2,k_-^2)}{(k_+^2-\MW^2)(k_-^2-\MW^2)}
        }_{\mbox{doubly-resonant}}
+\underbrace{\frac{R_{+}(k_+^2,k_-^2)}{k_+^2-\MW^2}
            +\frac{R_{-}(k_+^2,k_-^2)}{k_-^2-\MW^2}
        }_{\mbox{singly-resonant}}
+\underbrace{N(k_+^2,k_-^2)}_{\mbox{non-resonant}}\!,
\eeq
as explained in the introduction. The DPA amounts to the replacement
\beq
{\cal M} \to
\frac{R_{+-}(\MW^2,\MW^2)}{(k_+^2-\MW^2+\ri\MW\GW)(k_-^2-\MW^2+\ri\MW\GW)}.
\eeq
{\sloppy
Note that the numerator $R_{+-}(k_+^2,k_-^2)$ is replaced by the 
gauge-independent residue \linebreak $R_{+-}(\MW^2,\MW^2)$ \cite{st91,ae94}.
}

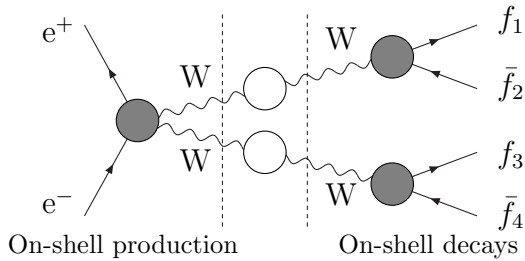
\begin{figure}
\centerline{
\begin{picture}(155,85)(0,0)
\SetScale{.8}
\ArrowLine(30,50)( 5, 95)
\ArrowLine( 5, 5)(30, 50)
\Photon(30,50)(150,80){2}{11}
\Photon(30,50)(150,20){2}{11}
\ArrowLine(150,80)(190, 95)
\ArrowLine(190,65)(150,80)
\ArrowLine(190, 5)(150,20)
\ArrowLine(150,20)(190,35)
\GCirc(30,50){10}{.5}
\GCirc(90,65){10}{1}
\GCirc(90,35){10}{1}
\GCirc(150,80){10}{.5}
\GCirc(150,20){10}{.5}
\DashLine( 70,0)( 70,100){2}
\DashLine(110,0)(110,100){2}
\put(40,21){W}
\put(40,53){W}
\put(95, 8){W}
\put(95,67){W}
\put(-12, 5){$\Pem$}
\put(-12,70){$\Pep$}
\put(160, 1){$\bar f_4$}
\put(160,24){$f_3$}
\put(160,50){$\bar f_2$}
\put(160,75){$f_1$}
\put(-25,-10){\footnotesize On-shell production}
\put(100,-10){\footnotesize On-shell decays}
\SetScale{1}
\end{picture}
}
\caption{Diagrammatic structure of factorizable corrections to
$\Pep\Pem\to\PW\PW\to 4f$}
\label{fig:fRCsdiag}
\end{figure}

\begin{figure}
\centerline{
\begin{picture}(110,75)(0,8)
\SetScale{0.8}
\ArrowLine(30,50)( 5, 95)
\ArrowLine( 5, 5)(30, 50)
\Photon(30,50)(90,80){2}{6}
\Photon(30,50)(90,20){2}{6}
\GCirc(30,50){10}{0}
\Vertex(90,80){1.2}
\Vertex(90,20){1.2}
\ArrowLine(90,80)(120, 95)
\ArrowLine(120,65)(105,72.5)
\ArrowLine(105,72.5)(90,80)
\Vertex(105,72.5){1.2}
\ArrowLine(120, 5)( 90,20)
\ArrowLine( 90,20)(105,27.5)
\ArrowLine(105,27.5)(120,35)
\Vertex(105,27.5){1.2}
\Photon(105,27.5)(105,72.5){2}{4.5}
\put(89,40){$\gamma$}
\put(42,60){$W$}
\put(42,15){$W$}
\SetScale{1}
\end{picture}
\begin{picture}(190,75)(0,8)
\SetScale{.8}
\ArrowLine(30,50)( 5, 95)
\ArrowLine( 5, 5)(30, 50)
\Photon(30,50)(90,80){2}{6}
\Photon(30,50)(90,20){2}{6}
\GCirc(30,50){10}{0}
\Vertex(90,80){1.2}
\Vertex(90,20){1.2}
\ArrowLine(90,80)(120, 95)
\ArrowLine(120,65)(105,72.5)
\ArrowLine(105,72.5)(90,80)
\ArrowLine(120, 5)( 90,20)
\ArrowLine( 90,20)(120,35)
\Vertex(105,72.5){1.2}
\PhotonArc(120,65)(15,150,270){2}{3}
\put(42,60){W}
\put(42,15){W}
\put(75,40){$\gamma$}
\DashLine(120,0)(120,100){6}
\PhotonArc(120,35)(15,-30,90){2}{3}
\Vertex(135,27.5){1.2}
\ArrowLine(150,80)(120,95)
\ArrowLine(120,65)(150,80)
\ArrowLine(120, 5)(150,20)
\ArrowLine(150,20)(135,27.5)
\ArrowLine(135,27.5)(120,35)
\Vertex(150,80){1.2}
\Vertex(150,20){1.2}
\Photon(210,50)(150,80){2}{6}
\Photon(210,50)(150,20){2}{6}
\ArrowLine(210,50)(235,95)
\ArrowLine(235, 5)(210,50)
\GCirc(210,50){10}{0}
\put(140,60){W}
\put(140,15){W}
\SetScale{1}
\end{picture}
}
\caption{Examples of virtual and real non-factorizable 
corrections to $\Pep\Pem\to\PW\PW\to 4f$}
\label{fig:nfRCsdiags}
\end{figure}
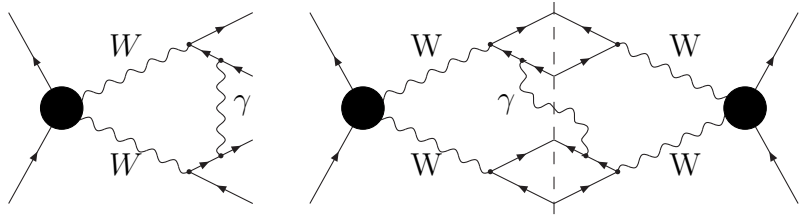

Doubly-resonant corrections to 
$\Pep\Pem\to\PW\PW\to 4f$ can be classified into two types
\cite{lep2repWcs,ae94,be94}:
factorizable and non-factorizable corrections.
The former are those that correspond either to W-pair production 
or to W~decay. They are represented by the schematic diagram of 
\reffi{fig:fRCsdiag}, in which the shaded blobs contain all one-loop
corrections to the production and decay processes, and the open blobs
include the corrections to the W~propagators. 
The remaining corrections
are called non-factorizable, since they do not contain the product of
two independent Breit--Wigner-type resonances for the W~bosons, i.e.\
the production and decay subprocesses are not independent in this case. 
Non-factorizable corrections include all diagrams involving particle
exchange between these subprocesses. Simple power-counting arguments
reveal that such diagrams only lead to doubly-resonant contributions if
the exchanged particle is a photon with energy
$E_\gamma\lsim\Gamma_\PW$; all other non-factorizable diagrams 
are negligible in DPA. Two relevant diagrams are shown in
\reffi{fig:nfRCsdiags}, where the full blobs represent tree-level
subgraphs.
We note that diagrams involving photon exchange between the W~bosons
contribute both to factorizable and
non-factorizable corrections; otherwise the splitting into those parts
is not gauge-invariant.
The non-factorizable corrections to $\Pep\Pem\to\PW\PW\to 4f$ 
are discussed in \refse{se:nfRCs} in more detail.

The factorizable corrections consist of contributions from virtual 
corrections and real-photon bremsstrahlung. The known results on
the virtual corrections to the pair production \cite{vrceeww}
and the decay \cite{rcwdecay} of on-shell W~bosons can be used as building
blocks for the DPA. 
The formulation of a consistent DPA for the 
real
corrections is, however, non-trivial. The main complication originates from
the emission of photons from the resonant W~bosons. 
A diagram with a radiating W~boson involves two propagators 
the momenta of which differ by the
momentum of the emitted photon. If the photon momentum is large 
($E_\gamma\gg\Gamma_\PW$), the resonances of these two propagators are
well separated in phase space, and their contributions can be associated
with photon radiation from exactly one of the production or decay
subprocesses. For soft photons ($E_\gamma\ll\Gamma_\PW$) a similar
splitting is possible. However, for $E_\gamma\sim\Gamma_\PW$ the two
resonance factors for the radiating W~boson overlap so that a simple
decomposition into contributions associated with the subprocesses is not
obvious.
\looseness -1

\subsection{Non-factorizable corrections}
\label{se:nfRCs}

Non-factorizable corrections 
account for the exchange of photons with $E_\gamma\lsim\Gamma_\PW$
between the W-pair production and W~decay subprocesses
(see \reffi{fig:nfRCsdiags}).
Already before their explicit calculation, it was shown
\cite{fa94} that such corrections 
vanish if the invariant masses of both W~bosons are integrated over. 
Thus, they do not influence pure angular distributions,
which are of particular importance for the analysis of gauge-boson
couplings. For exclusive quantities the non-factorizable corrections are 
non-vanishing. A first hint on their actual size was obtained by 
investigating the non-factorizable correction that is contained in the 
Coulomb singularity \cite{kh95}.

The explicit analytical calculation of the non-factorizable corrections was 
performed by different groups \cite{me96,be97,de98a}%
\footnote{The original result of the older calculation \cite{me96} does
not agree with the two more recent results \cite{be97,de98a}, which are
in mutual agreement. As known 
from the authors of \citere{me96}, 
their corrected results also agree with the ones of \citeres{be97,de98a}.}.
In these studies, the photon momentum was integrated over, resulting in
a correction factor to the differential Born cross section for the process
without photon emission. This correction factor is non-universal \cite{de98a} 
in the sense that it depends on the
parametrization of phase space. 
The analytical results show that
all effects from the initial $\Pep\Pem$ state
cancel so that the correction factor does not depend on the
W-production angle.
Fermion-mass singularities appear in
individual contributions, but cancel in the sum. 
Moreover, the correction factor vanishes
like $(M_\pm^2-\MW^2)/(\Gamma_\PW\MW)$ on resonance and tends to zero in
the high-energy limit, both leading to a suppression of the non-factorizable
corrections with respect to the factorizable ones.

\begin{figure}
\centerline{
\setlength{\unitlength}{1cm}
\begin{picture}(11,5)
\put(0,0){\includegraphics{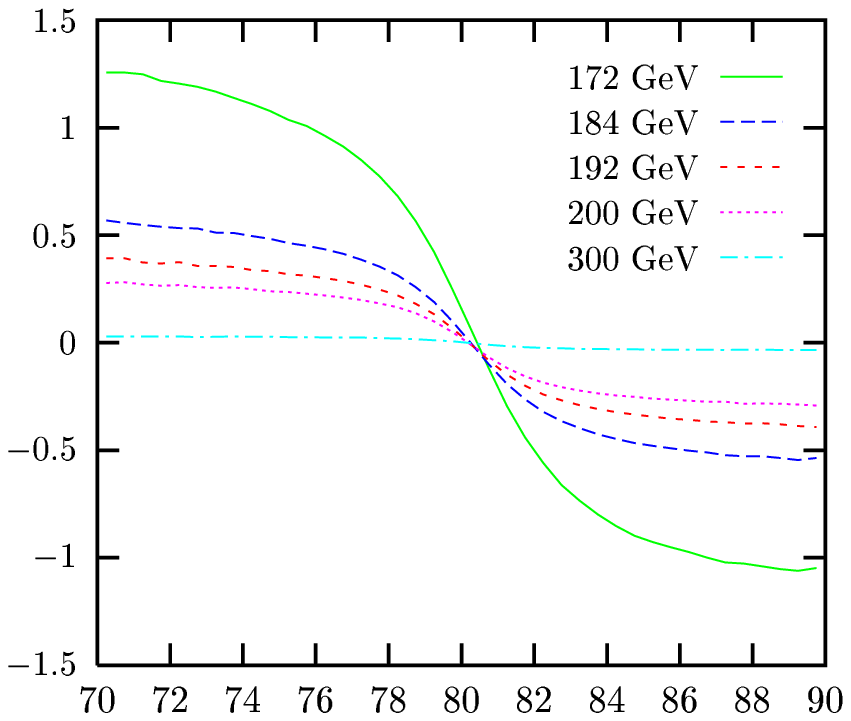}}
\put(0,0){\includegraphics{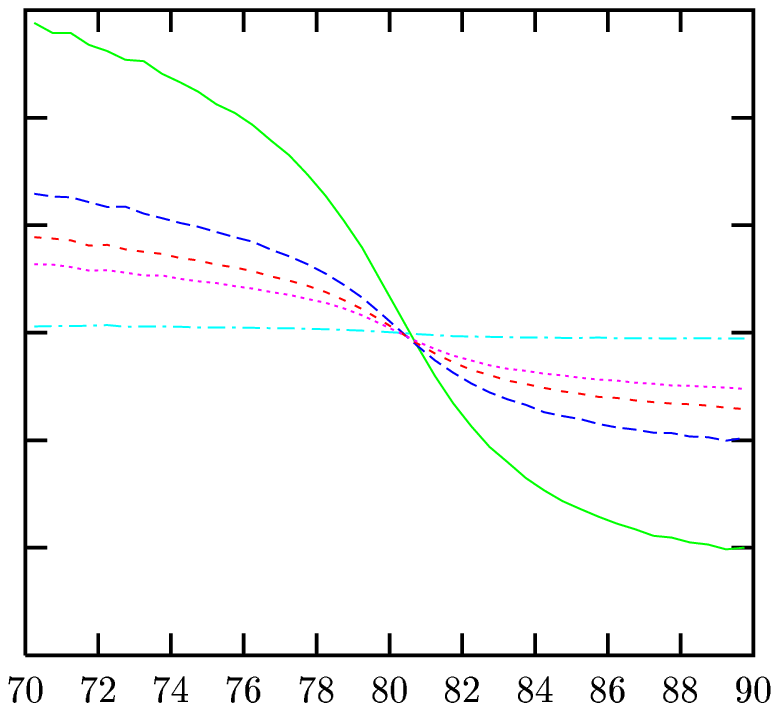}}
\put(-0.4,3.9){\makebox(1,1)[c]{\small$\de_{\mathrm{nf}}/\%$}}
\put(2.5,-0.6){\makebox(1,1)[cc]{{\small$M_\pm/{\GeV}$}}}
\put(7.8,-0.6){\makebox(1,1)[cc]{{\small$M_\pm/{\GeV}$}}}
\put(1.4,1.1){\footnotesize leptons}
\put(6.7,1.1){\footnotesize hadrons}
\end{picture}
}
\caption{Relative non-factorizable corrections to single-invariant-mass
distributions for $\Pep\Pem\to\PW\PW\to 4f$ with purely leptonic 
and hadronic final states (taken from \protect\citere{de98a})}
\label{fig:nonfactnum}
\end{figure}

The non-factorizable corrections to $\Pep\Pem\to\PW\PW\to 4\,$leptons were 
numerically evaluated in \citere{be97} and for all final states in 
\citere{de98a}.
The corrections to 
invariant-mass distributions 
(see \reffi{fig:nonfactnum}) turn out to be
qualitatively similar for all final states and are of the order
of $\sim 1\%$ for LEP2 energies, shifting the maximum of the distributions
by 1--2$\MeV$.
Multiple distributions in angular or energy variables and in at least
one of the invariant masses of the W~bosons receive larger corrections
of a few per cent.

Although non-factorizable corrections to four-fermion production turn out 
to be small with respect to LEP2 accuracy, they
can be of relevance at future $\Pep\Pem$ colliders with higher
luminosity.

\subsection{Results for ${\cal O}(\alpha)$ corrections in
double-pole approximation}

In \citere{be98b} the ${\cal O}(\alpha)$ corrections to four-lepton 
production were treated in DPA, following a semi-analytical approach.
The DPA is applied both to the virtual and real corrections, and the
off-shellness of the W~bosons 
is kept only in the W~propagators, but
nowhere else. In particular, the phase space is factorized into on-shell 
phase spaces and independent invariant masses $M_\pm$ for the W~bosons. 
The corresponding W-boson momenta $k_\pm$ were strictly identified with 
the sum of the momenta of the corresponding decay fermions, i.e.\ no photon
recombination was considered. For the total cross section 
the approach of \citere{be98b}
is closely related to taking the cross section for on-shell W-pair 
production multiplied by a branching ratio. \citere{be98b} also contains
results for the $M_\pm$ distributions and various angular
distributions for a CM energy of $184\GeV$. In particular, the authors
of \citere{be98b} find relatively large shifts in the peak position of
the W~line shape, namely $-20\MeV$, $-39\MeV$, and $-77\MeV$ for
$\tau^+\nu_\tau$, $\mu^+\nu_\mu$, and $\Pep\nu_\Pe$ final states
respectively. These results have been qualitatively confirmed by YFSWW
in \citere{ja99}, where the ${\cal O}(\alpha)$ corrections to W-pair
production \cite{ja97} were supplemented by final-state radiation in a 
leading-log approach. 
Note, however, that the large shifts are due to mass-singular
logarithms like $\alpha\ln(m_l/\MW)$, since no photon recombination of
collinear photons is performed. More realistic
definitions of $k_\pm^2$, which have to include photon recombination,
effectively replace the mass-singular logarithms by logarithms of a
minimum opening angle for collinear photon emission. This expectation is
also confirmed by the leading-log study of \citere{ja99}.

Moreover, a DPA for the factorizable corrections to 
$\Pep\Pem\to\PW\PW\to\Pu\bar\Pd\bar\Pc\Ps$ was used in \citere{ku99} in
order to estimate the quality of a high-energy approximation
\cite{be93,ku98} for the virtual corrections at energies 
$\sqrt{s}\gsim 500\GeV$. \citere{ku99} contains results on total cross
sections and distributions in the W~production angle.

Very recently, we succeeded in constructing the first Monte
Carlo generator for off-shell \PW-pair production that includes the
complete $\Oa$ corrections in DPA. 
This generator, called {\sc RacoonWW}, contains
the full lowest-order matrix elements for $\eeffff$ for any
four-fermion final state. The complete virtual corrections to \PW-pair
production and \PW~decay, and the virtual non-factorizable corrections
are included in the DPA.  The exact four-fermion phase space is used
throughout. For the real corrections the matrix elements for the
minimal gauge-invariant subset comprising all doubly-resonant
contributions of the processes $\eeWWffffg$, \ie the photon radiation
from the CC11 subset, are included. By using these matrix elements for
the real radiation, we avoid the problems in defining a DPA for
semi-soft photons ($E_\gamma\sim\GW$). 
The real and virtual corrections are carefully combined in the soft and 
collinear regions, in order to avoid mismatch between IR and mass 
singularities.
\looseness -1

\begin{figure}
\centerline{
\setlength{\unitlength}{1cm}
\begin{picture}(6,5)
\put(0,0){\includegraphics{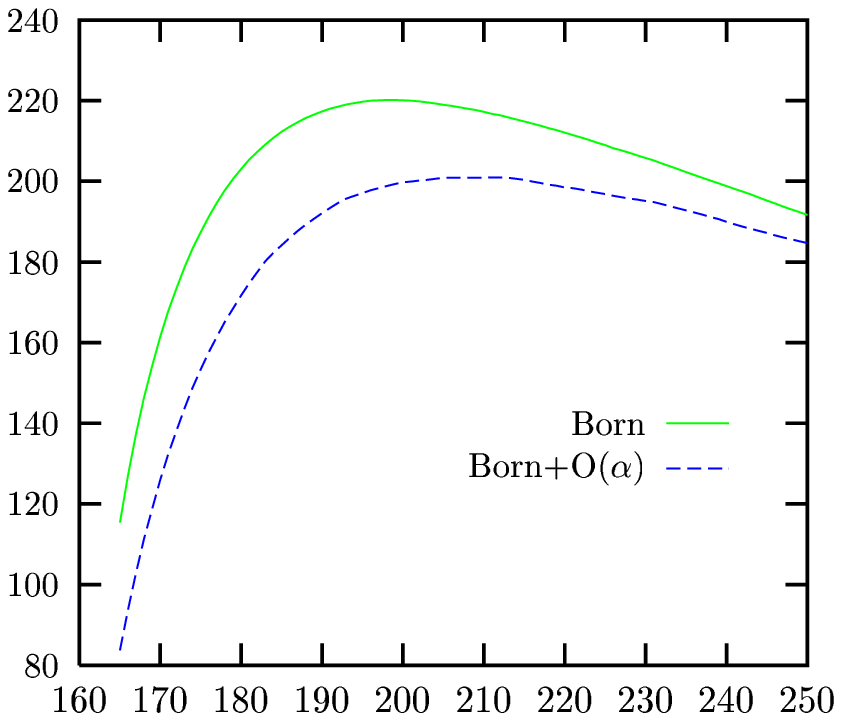}}
\put(2.8,-0.6){\makebox(1,1)[cc]{{\small $\sqrt{s}/\GeV$}}}
\end{picture}
\begin{picture}(6,5)
\put(0,0){\includegraphics{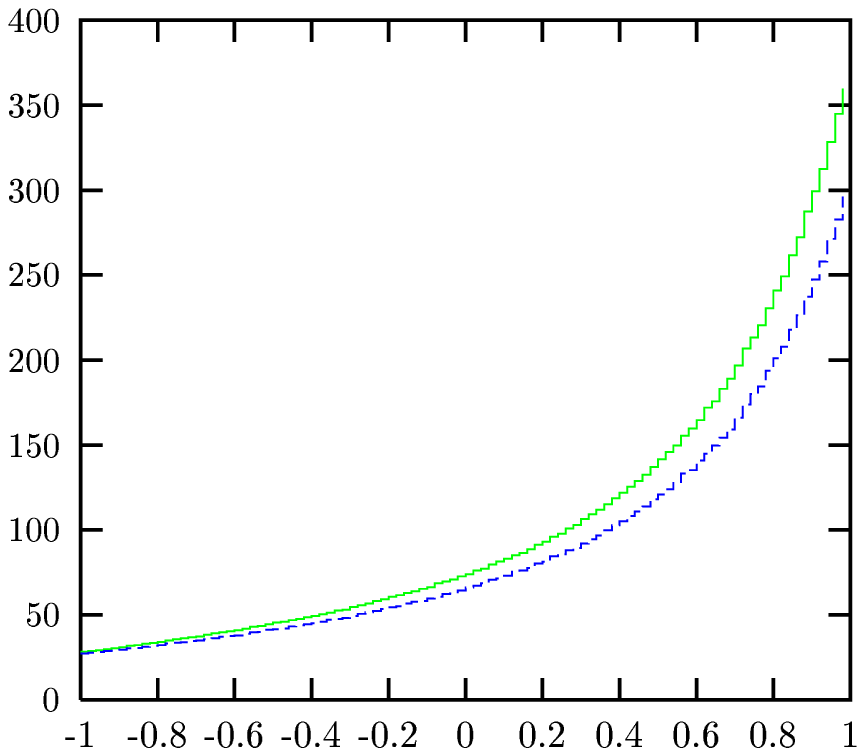}}
\put(2.5,-0.6){\makebox(1,1)[cc]{{\small $\cos\theta$}}}
\put(2.0,3.5){\makebox(1,1)[cc]{{\small $\sqrt{s} = 184\GeV$}}}
\end{picture} }
\label{fi:fullrcs}
\caption[]{Total cross section $\sigma$ (left) and differential
cross section $\rd\sigma/\rd\cos\theta$ (right), both in fb, 
for $\Pep\Pem\to\PW\PW\to\tau^+\nu_\tau\mu^-\bar\nu_\mu$}
\end{figure}

As a first result of this generator, 
we show the ${\cal O}(\alpha)$-corrected total cross section and the
distribution in the W-production angle $\theta$ for four-lepton
production in \reffi{fi:fullrcs}.
A detailed presentation of
numerical results as well as a comparison to existing results will
appear elsewhere.

\frenchspacing
\newcommand{\app}[3]{{\sl Acta Phys.\ Pol.} {\bf #1} (19#2) #3}
\newcommand{\ap}[3]{{\sl Ann.~Phys.} {\bf #1} (19#2) #3}
\newcommand{\zp}[3]{{\sl Z.~Phys.} {\bf #1} (19#2) #3}
\newcommand{\np}[3]{{\sl Nucl.~Phys.} {\bf #1} (19#2) #3}
\newcommand{\pl}[3]{{\sl Phys.~Lett.} {\bf #1} (19#2) #3}
\newcommand{\prep}[3]{{\sl Phys.\ Rep.} {\bf #1} (19#2) #3}
\newcommand{\pr}[3]{{\sl Phys.~Rev.} {\bf #1} (19#2) #3}
\newcommand{\prl}[3]{{\sl Phys.~Rev.~Lett.} {\bf #1} (19#2) #3}
\newcommand{\fp}[3]{{\sl Fortschr.~Phys.} {\bf #1} (19#2) #3}
\newcommand{\jp}[3]{{\sl J.~Phys.} {\bf #1} (19#2) #3}
\newcommand{\cpc}[3]{{\sl Comput.~Phys.~Commun.} {\bf #1} (19#2) #3}
\newcommand{\ijmp}[3]{{\sl Int.~J.~Mod.~Phys.} {\bf #1} (19#2) #3}
\newcommand{\nim}[3]{{\sl Nucl.~Instr.~Meth.} {\bf #1} (19#2) #3}
\newcommand{\nc}[3]{{\sl Nuovo Cimento} {\bf #1} (19#2) #3}

\newcommand{\vj}[4]{{\sl #1} {\bf #2} (19#3) #4}

\end{document}